\documentclass{article}
\usepackage{arxiv}
\usepackage[utf8]{inputenc} 
\usepackage[T1]{fontenc}    
\usepackage{hyperref}       
\usepackage{url}            
\usepackage{booktabs}       
\usepackage{amsfonts}       
\usepackage{nicefrac}       
\usepackage{microtype}      
\usepackage{lipsum}
\usepackage{graphicx}
\usepackage{xcolor}

\title{Segmentation with Residual Attention U-Net\\ and an Edge-Enhancement Approach\\ Preserves Cell Shape Features}

\author{
 Nanyan Zhu\thanks{These authors contributed equally.}\\
  Department of Biological Sciences\\
  Columbia University\\
  New York, NY, 10027\\
   \And
 Chen Liu\footnotemark[1] \\
  Department of Electrical Engineering\\
  Columbia University\\
  New York, NY, 10027\\
  \And
 Zakary S. Singer\\
  Department of Biomedical Engineering\\
  Columbia University\\
  New York, NY, 10027\\
  \And
 Tal Danino\\
  Department of Biomedical Engineering\\
  Columbia University\\
  New York, NY, 10027\\
  \And
 Andrew F. Laine\thanks{Correspondence: Jia Guo (jg3400@columbia.edu) or Andrew F. Laine (al418@columbia.edu). \newline \textcolor{white}{......}The project is accessible via the GitHub repository \url{https://github.com/SAIL-GuoLab/Cell_Segmentation_and_Tracking}.}\\
  Department of Biomedical Engineering\\
  Columbia University\\
  New York, NY, 10027\\
  \And
 Jia Guo\footnotemark[2]\\
  Department of Psychiatry, and\\
  Mortimer B. Zuckerman Mind Brain Behavior Institute\\
  Columbia University\\
  New York, NY, 10027\\
}

\begin{document}
\maketitle
\begin{abstract}
The ability to extrapolate gene expression dynamics in living single cells requires robust cell segmentation, and one of the challenges is the amorphous or irregularly shaped cell boundaries. To address this issue, we modified the U-Net architecture to segment cells in fluorescence widefield microscopy images and quantitatively evaluated its performance. We also proposed a novel loss function approach that emphasizes the segmentation accuracy on cell boundaries and encourages shape feature preservation. With a 97\% sensitivity, 93\% specificity, 91\% Jaccard similarity, and 95\% Dice coefficient, our proposed method called Residual Attention U-Net with edge-enhancement surpassed the state-of-the-art U-Net in segmentation performance as evaluated by the traditional metrics. More remarkably, the same proposed candidate also performed the best in terms of the preservation of valuable shape features, namely area, eccentricity, major axis length, solidity and orientation. These improvements on shape feature preservation can serve as useful assets for downstream cell tracking and quantification of changes in cell statistics or features over time.
\end{abstract}

\section{Introduction}
\label{sec:intro}

Observations of cell dynamics at the population level is inherently sub-optimal due to the heterogeneity among individual cells. In contrast, imaging approaches enable single cell analysis, but require robust delineation of cell boundaries (aka., cell segmentation) and, for temporal information, matching cell identities over time (aka., cell tracking). Simpler cases with desirable properties such as high foreground-background contrast, high signal-to-noise ratios, and similar cell appearances, can be well handled by automated algorithms. On the other hand, manual annotation by experts is necessary whenever complications arise. Since the manual task of labeling and tracking is both tedious and may be inefficient and inconsistent when annotated by different individuals, a robust and automated alternative is highly desired.

The most promising work comes from the field of computer vision, where the topics of instance segmentation (i.e. treating multiple objects of the same category as separable individuals, as opposed to semantic segmentation) and multi-object tracking have been heavily visited. As summarized in Ciaparrone et al \cite{DL_MOT_review}, segmentation and tracking tasks are typically addressed with four sequential steps: object detection, feature extraction, similarity calculation and identity association. While a lot of studies has been place on each of these four steps, the necessity of preserving cell shape features is underrepresented in the literature. Additionally, traditional evaluation metrics to quantify segmentation performance, such as accuracy, specificity, and Jaccard similarity, do not directly reflect how well a cell boundary is delineated. For example, if a star-shaped cell whose long processes constitutes 3\% of its total area is approximated with a simple elliptical segmentation, the scores given by the aforementioned metrics can go beyond 95\%, even though essential shape features are completely discarded.

Morphological details carry important information about the cell. For instance, the various unusual shapes of red blood cells reflect different developmental stages and/or pathologies \cite{Red_cell_shape}. Efforts spent on preserving cell shape features not only benefit the performance of various tracking algorithms that primarily rely on shape features such as Dewan et al \cite{Cell_track_feature}, but also enable more accurate quantification of cell characteristics since the segmentation result is more reliable.

\section{Related Work}
\label{sec:relatedwork}
Computational techniques adopted for cell segmentation range from thresholding \cite{Cell_seg_thresholding_canny} on the simpler end along the spectrum of methods to deep learning and graph theory \cite{cell_joint_seg_track_graphical} toward the more sophisticated end. The method should be selected on a case-specific basis, i.e., a straightforward and effective algorithm for one case may completely fail for a second, while a rigorous and robust algorithm for the second may be overly complicated for another, etc. For instance, Al-Hafiz et al \cite{Cell_seg_thresholding_canny} could achieve an 87.9\% accuracy rate by merely applying a dynamic thresholding algorithm on the image gradient and refining with morphological operations because their red blood cells are mostly homogeneous in intensity, regularly shaped, and clearly distinguishable from the background. Bensch et al \cite{Cell_seg_track_asymmetric_boundary_cost}, on the other hand, needed to use a more advanced technique that utilized graph cut along with an asymmetric cost function over the edges to achieve better boundary delineation due to the complication of noisy acquisition, shading effects, and irregular cell morphology.

A few benchmarks for cell segmentation, or instance segmentation in general, include the U-Net family \cite{U-Net} and the R-CNN family \cite{RCNN} of neural networks. The former is known for wide application in biomedical image segmentation while the latter excels at proposing likely objects and evaluating such detection proposals.

Datasets complicated by bias fields, visual artifacts, ambiguous cell boundaries, irregular cell shapes, overlapping or touching cells and other factors collectively render the problem significantly more challenging. One of the best performing algorithms in a similarly complicated dataset is introduced by Falk's team \cite{U-Net_cell_seg}, where they applied the aforementioned U-Net on cell segmentation and yielded segmentation results comparable to human experts. Therefore we implemented the same deep learning architectures as they did and made performance-boosting adjustments on a variety of aspects from architecture and loss function to image processing techniques.

\section{Materials and Methods}
\label{sec:materialsandmethods}
\subsection{Dataset}
\begin{figure}[t!]
\centering
\includegraphics[width = 420pt]{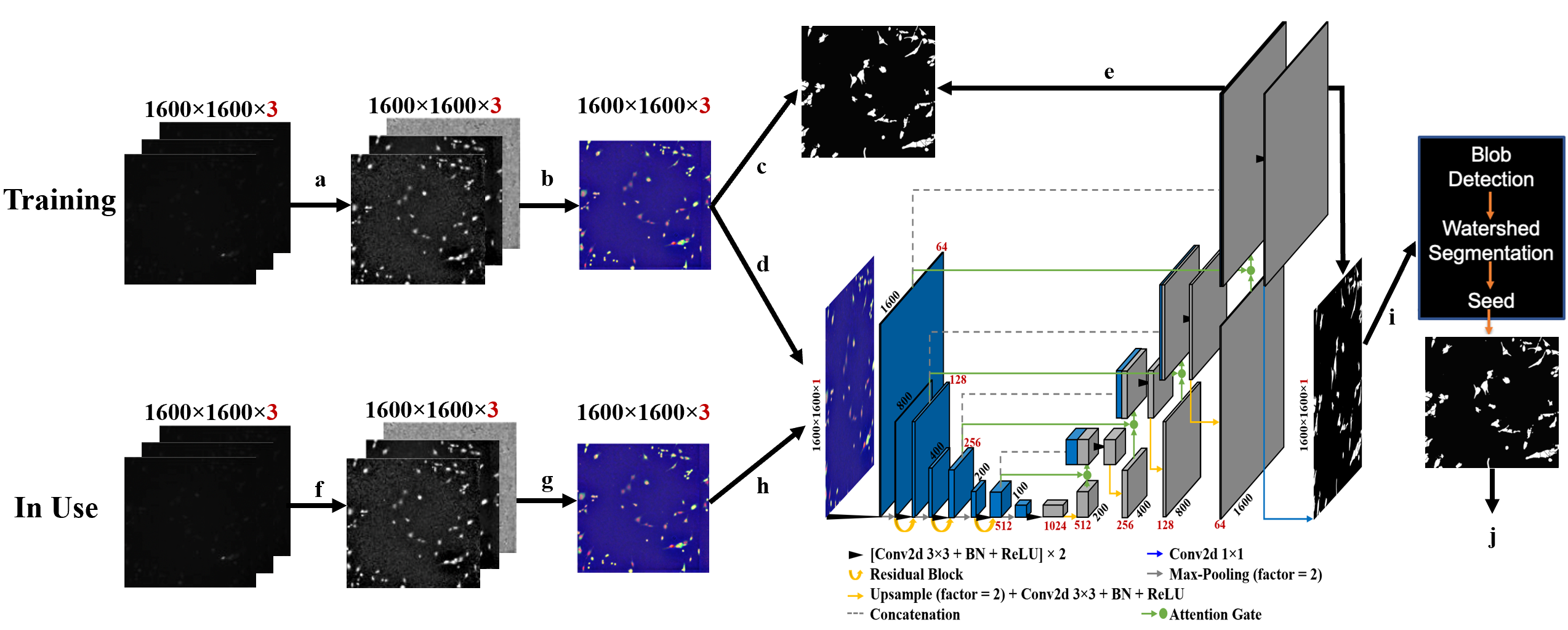}
\caption{\textbf{Demonstration of the segmentation and tracking pipeline.} \textbf{a.\&f.} Bias field correction and intensity normalization. \textbf{b.\&g.} Channel-combination to form pseudo-RGB images. \textbf{c.} Manual annotation. \textbf{d.\&h.} Provide as input. \textbf{e.} Use the manual annotation as the ground truth to tune the deep learning network. Only provided during training and validation but not during testing. \textbf{i.} Post-processing. \textbf{j.} Future tracking algorithms.}
\label{Seg_Track_Pipeline}
\end{figure}

The data we used was acquired at Columbia University in the department of Biomedical Engineering, where images were collected on the same baby hamster kidney (BHK) cell cultures every 20 minutes over roughly 15 hours. These images were acquired using fluorescence widefield microscopy images of nuclear and cytoplasmic live cell stains, differential interference contrast (DIC), and fluorescent proteins from cytoplasmically localized reporters. How the data is subsequently processed is described in the upcoming subsections as well as illustrated in Figure \ref{Seg_Track_Pipeline}.

\subsection{Pre-processing and Data Augmentation}
The nuclear and cytoplasmic stains, along with the DIC images, are treated as three channels of information for the localization of cells, while additional fluorescent reporters are treated as cell parameters that can be extracted once the segmentation and tracking is complete. The three channels are first bias-field corrected and intensity normalized by dividing each of them by its respective Gaussian filtered version. A histogram-based intensity normalization is then performed on each of the three images such that a selected percentage of the brightest and darkest intensities are clipped while the remaining values are linearly mapped to the range of [0, 1]. Next, the 3 channels are concatenated to form pseudo-RGB images (red: nuclei, green: cytoplasm, blue: DIC). Manual annotation of the cells are done on the pseudo-RGB images, and the pseudo-RGB are fed into the deep learning network as the input representations with the manual annotations as the ground truths that guide the training process.

Data augmentation is done by random horizontal and vertical flipping, as well as random patch selection on the 4,792$\times$3,200 regions yielding patches of dimension 1,600$\times$1,600. Scaling and affine deformations are not considered as reasonable options since they may alter the naturally-occurring shape features. Also the data augmentation is only implemented for the training and validation sets. In total, we used approximately 4,600 cells in the training set, 1,000 in the validation set and 500 in the test set.

\subsection{Deep Learning Architecture for Segmentation}
The deep learning architecture adopted is a five-layered Residual Attention U-Net (ResAttU-Net), partially inspired by an open-source GitHub repository \cite{pytorch_unet_family}. The major differences between this architecture and the popular U-Net is the addition of two pieces in the former: Residual Blocks introduced in He at el \cite{Residual_block} and Attention Mechanism introduced in Oktay et al \cite{AttU-Net}.

\subsection{The Edge-Enhancement Approach}
\begin{figure}[t!]
\centering
\includegraphics[width = 450pt]{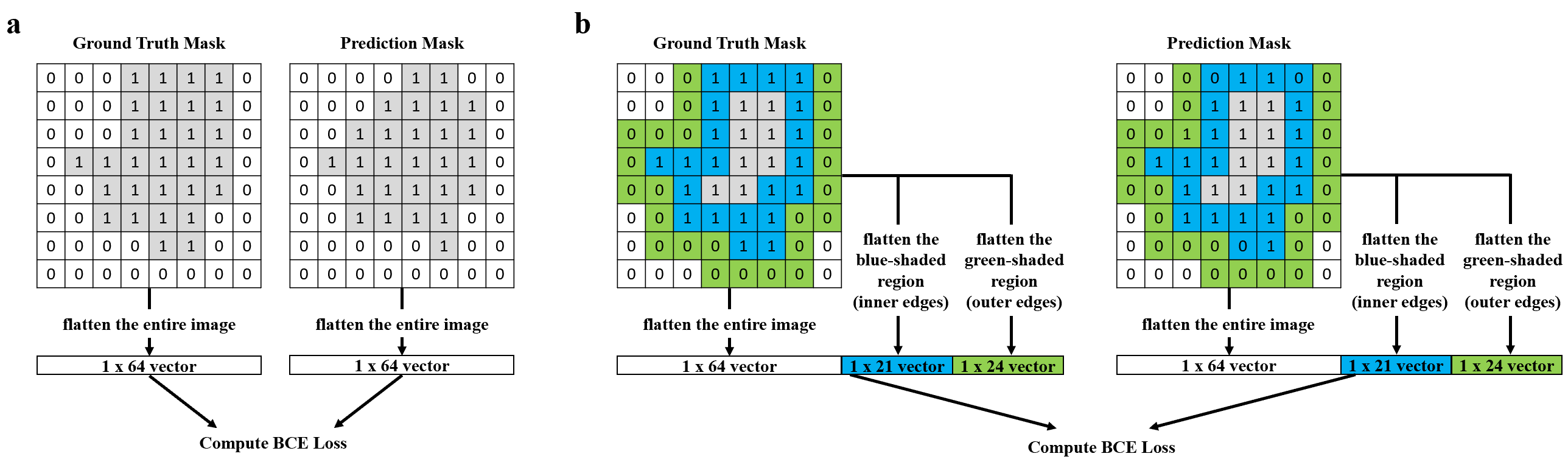}
\caption{\textbf{Illustration of the Edge-Enhancement Approach on an Example Patch of Dimension 8$\times$8.} \textbf{a.} How binary cross entropy loss is typically calculated in a segmentation task. \textbf{b.} How we propose to incorporate the edge information twice in the calculation so that the accurate prediction of the edges is emphasized.}
\label{Edge_Enhancement}
\end{figure}

Due to the desire to retain the long cellular extensions characteristic to our cell-line of interest, BHK-21, a special mechanism we implemented is a novel loss function approach that we call edge-enhancement (EE). Normally the loss function is defined as the binary cross-entropy loss between the deep-learning-predicted segmentation and the manually-annotated segmentation ground truth after flattening them into vectors (Figure \ref{Edge_Enhancement}a). For edge-enhancement, we deliberately emphasize the weighting on the accurate prediction of the cell edges (Figure \ref{Edge_Enhancement}b). Specifically, we find the Laplacian-of-Gaussian of the ground truth segmentation, take two binary-thresholded versions of the resulting image (intensities greater than 0.001 for one and intensities less than -0.001 for the other) that respectively corresponds to the inner and outer cell boundaries, vectorize the foreground regions as well as their corresponding regions in the deep learning prediction, and append them to their respective segmentation vectors for loss calculation.

\subsection{Post-processing of the Segmented Cells}
Once the deep learning cell segmentation is generated, it is then refined by watershed splitting of adjacent touching cells, where the seeds for watershed are extracted using blob detection on the intensity-rescaled difference image between the nuclei image and the cytoplasm image. The cleaned up segmentation result in the form of instantiated cell masks, can be directly provided as input to any suitable downstream tracking algorithm.

\section{Results}
\label{sec:results}
\subsection{Evaluation of Cell Segmentation: Traditional Metrics}
The segmentation performance of the candidates being compared are listed in Table \ref{Seg_Result_Table}. Among all candidates, ResAttU-Net with our edge-enhancement approach achieved the best performance in all metrics evaluated. With 97\% sensitivity and 93\% specificity, the method we propose outperformed a segmentation benchmark, U-Net \cite{U-Net_cell_seg}.

\begin{table}[ht!]
\centering
\caption{\textbf{Evaluation of the segmentation using traditional quantitative metrics.}}
\includegraphics[width = 350pt]{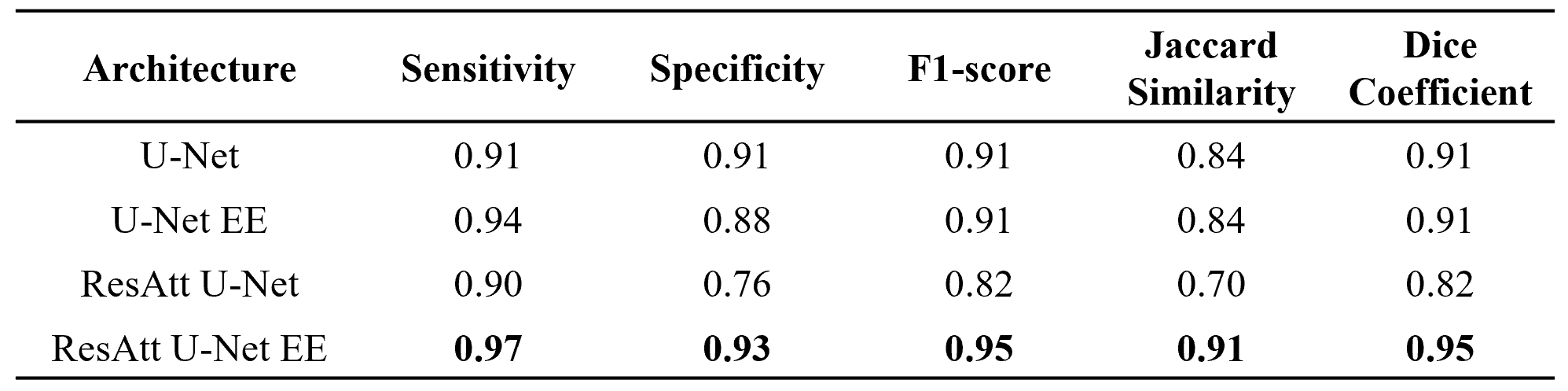}
\label{Seg_Result_Table}
\end{table}

\begin{figure*}[t!]
\centering
\includegraphics[width = 400pt]{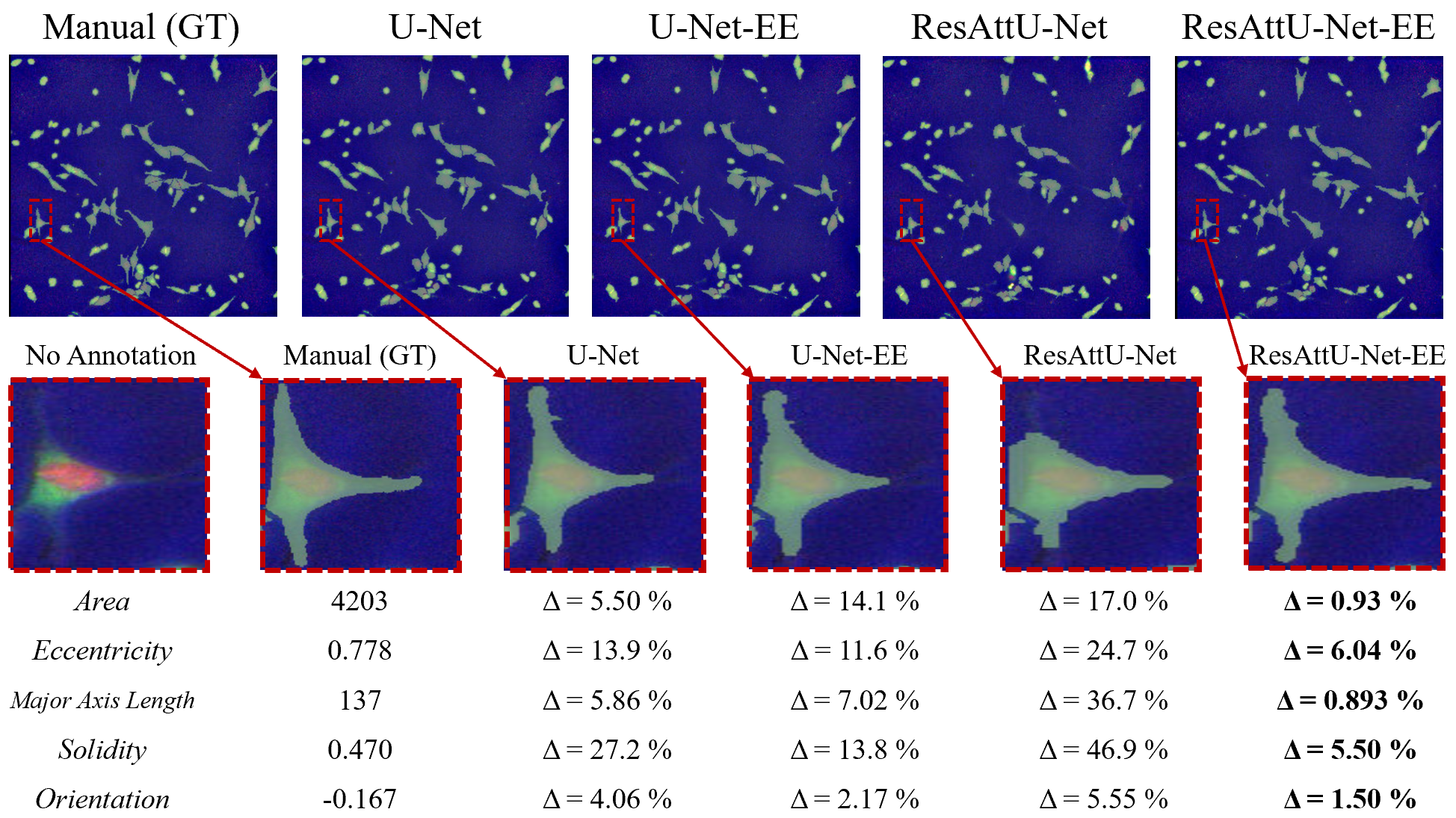}
\caption{\textbf{Visualization of the segmentation and demonstration of the cell features.} The original pseudo-RGB images (R: nuclei, G: cytoplasm, B: DIC) are overlayed with the respective segmentation masks in light green. While the overall segmentation quality are comparable among all four candidates, it can be seen that the cytoplasm extensions are best captured by the ResAttU-Net with edge enhancement (see Section 3.4). The percentage change of each shape feature with respect to the manual ground truth also indicates a better shape feature preservation in that candidate. Note that for orientation, 100 percent difference represents a mismatch of 90 degrees.}
\label{Seg_result}
\end{figure*}

\subsection{Evaluation of Cell Segmentation: Shape Feature Preservation}
\begin{figure*}[ht!]
\centering
\includegraphics[width = 450pt]{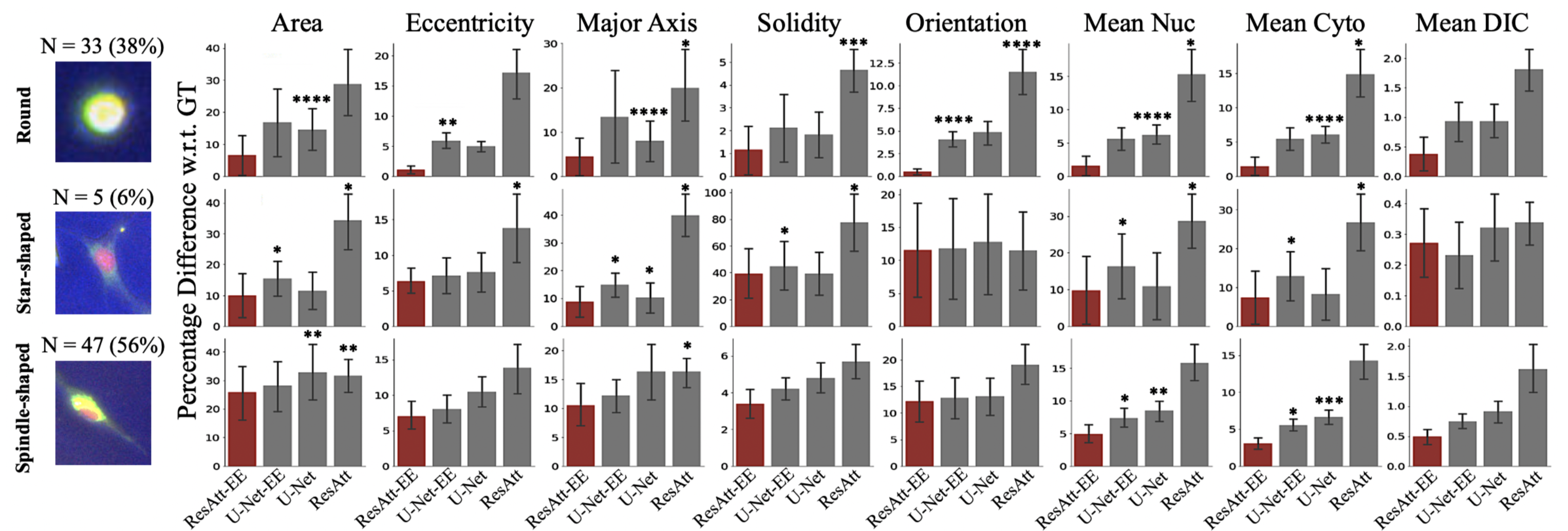}
\caption{\textbf{Evaluation of the segmentation using shape feature preservation metrics.} For each category of cell shape, an example image is displayed and the percentage change with respect to the ground truth are reported for all four segmentation candidates. The bar graphs indicate the Mean $\pm$ Standard Error of the Mean (SEM). One-tailed paired t-tests are performed with respect to the best-performing Residual Attention U-Net with edge-enhancement, and the significance levels are represented as asterisks (1, 2, 3, 4 asterisks respectively corresponds to p-values less than 0.05, 0.01, 0.001, 0.0001). The acronyms are modified in the figure such that ResAtt stands for ResAttU-Net and EE stands for edge-enhancement.}
\label{Shape_Feature}
\end{figure*}

We first grouped a total number of 85 randomly selected cells from the test set into three shape categories based on their convexity, as calculated by the area of the cell divided by the area of its convex hull. Highly-convex cells (convexity $>$ 0.95) are categorized as round, least-convex cells (convexity $<$ 0.5) as star-shaped, while the ones in between as spindle-shaped. Afterwards, five shape features were extracted from the respective segmentation mask of each cell. From Figure \ref{Seg_result} and Figure \ref{Shape_Feature}, it can be seen that our proposed method performed the best in terms of cell shape feature preservation. As a result, the mean intensity values of the three information-providing channels within the segmentation mask given by our proposed method also best resembled those in the manual annotation ground truth masks. It is reasonable to believe that the same will hold true for other information to be extracted from these segmentation results.

\section{Discussions}
\label{ssec:discussions}

\subsection{Irreplaceability of Shape Feature Metrics}
Traditional evaluation metrics (Table \ref{Seg_Result_Table}) imply barely distinguishable performance of the two U-Net candidates regardless of edge-enhancement. However, the shape feature metrics (Figure \ref{Shape_Feature}) reveals their advantages and disadvantages in a more comprehensive manner. We can thus conclude that shape feature metrics provide insights that are not reflected in the traditional metrics.

\subsection{Effect of Edge-enhancement on U-Net}
The effect of edge-enhancement on U-Net is better explained by the shape features. We observed that the version with edge-enhancement performed slightly better for the spindle-shaped cells but slightly worse for cells in the other two categories. To our best understanding, this phenomenon might be attributed to the fact that cells of spindle shapes were more prevalent in the training data. It is possible that the U-Net does gain a better grasp of shape features under the help of edge-enhancement, but the improvement is somewhat coarse and is tailored better towards the more frequently occurring cell shapes. Since there are more spindle-shaped cells in the training data, the deep learning model, despite having difficulty learning the nuances of cells from all shape categories, is nevertheless able to provide a better segmentation for cells from the most prevalent shape category.

\subsection{Effect of Edge-enhancement on ResAttU-Net}
Unlike the more ambiguous case with U-Net, the edge-enhancement approach boosts the segmentation performance of ResAttU-Net on almost every aspect. We hypothesize that a network with the attention mechanism would benefit more from edge-enhancement. Indeed this was the case, since the attention mechanism enabled the network to better understand / encode the regional contribution to the loss, and therefore the network was able to know which parameters to adjust in response to the training feedback. In this way, it was better able to adjust itself to preserve the shape features for all cell shapes.

\subsection{Potential Utility of our Proposed Candidate}
The ability to identify and follow individual cells over time has provided insight into processes ranging from bacterial dynamics to mammalian development \cite{muzzey2009quantitative}. A central challenge across all of these applications is robust cell segmentation. This presents an especially difficult challenge when identifying the entirety of the cytoplasm of complex and motile cells who may have irregular shapes and/or long and thin cellular processes. This contrasts with simple alternative methods that rely exclusively on identifying the nucleus of a cell, but which necessarily exclude the study of cytoplasmic processes or fluorescent reporters not amenable to nuclear reporters. The method described here helps overcome the limitations of alternative segmentation approaches and preserves the unusual processes and extensions observed in certain primary cells or cell lines, and will likely find relevance beyond the BHK-21 line studied here.

\section{Acknowledgements}
\label{sec:acknowledgements}
This study was initiated and supported by Zakary S. Singer and Tal Danino from Columbia Biomedical Engineering. The computational resources are provided by Jia Guo and Tommy Vaughan from Mortimer B. Zuckerman Mind Brain Behavior Institute.

\bibliography{references.bib}
\bibliographystyle{unsrt}

\end{document}